\documentclass[12pt]{iopart}
\usepackage[dvips]{graphicx}
\usepackage{amsfonts}
\usepackage{color}
\usepackage{colortbl}
\usepackage{hyperref}
\usepackage[numbers,sort&compress]{natbib}
\usepackage{hypernat}

\begin{document}
\title[Patterns of cooperation]{Patterns of cooperation: fairness and coordination in networks of interacting agents}

\author{Anne-Ly Do, Lars Rudolf, and Thilo Gross}

\address{Max-Planck-Institute for the Physics of Complex Systems, Dresden, Germany}
\ead{ly@pks.mpg.de}

\begin{abstract}
We study the self-assembly of a complex network of collaborations among self-interested agents. The agents can maintain different levels of cooperation with different partners. Further, they continuously, selectively, and independently adapt the amount of resources allocated to each of their collaborations in order to maximize the obtained payoff. We show analytically that the system approaches a state in which the agents make identical investments, and links produce identical benefits. Despite this high degree of social coordination some agents manage to secure privileged topological positions in the network enabling them to extract high payoffs. Our analytical investigations provide a rationale for the emergence of unidirectional non-reciprocal collaborations and different responses to the withdrawal of a partner from an interaction that have been reported in the psychological literature.
\end{abstract}
\submitto{\NJP}
\maketitle

Cooperation is the basis for complex organizational structures in biological as well as in social systems \cite{AxelrodHamilton,DoebeliHauert}. 
The evolutionary and behavioural origin of cooperation is a subject of keen scientific interest, because the ubiquity of cooperation in nature seems to defy the often high costs incurred by the cooperating agent \cite{NowakSigmund}. 
Evolutionary game theory has identified several mechanism allowing for the evolution and persistence of costly cooperation \cite{Nowak}. 
In particular the emergence of cooperation is promoted if the interacting agents are distributed in some  (potentially abstract) space, so that only certain agents can interact at any given time \cite{Axelrod,NowakMay,Turchin}. 
In the context of social cooperation spatial structure can be appropriately modeled by a complex network, in which nodes represent agents, while the links correspond to collaborations. 
The topology of this network, i.e., the specific configuration of nodes and links, has been shown to be of central importance for the level of cooperation that evolves \cite{HauertDoebeli,EguiluzZimmermannCela,SantosPacheco,OhtsukiHauert,SantosSantos}.

In social networks the topology is not static, but reacts to the behaviour of the agents \cite{Macy,Gould,Willers,FehrFischbacher,Barabasi,Braha}. 
This defines an inherent dynamical interplay: While the agents's behaviour may depend on their topological neighbourhood, this neighbourhood is, at least in part, shaped through the agent's behavioural choices. 
Networks containing such an dynamical interplay between the state of the nodes and the networks topology 
are called \emph{adaptive networks} \cite{GrossBlasius,Gross2009}. While adaptive networks have been studied for some time in the social literature (e.g. \cite{Macy,Ashlock,BalaGoyal}), pioneering work \cite{BornholdtRohlf, Zimmermann2000, PemantleSkyrms, Paczuski2000} only recently 
triggered a wave \cite{Wiki} of detailed dynamical investigations in physics. 
Recent publications discuss simple cooperative games such as the one-shot prisoner's dilemma
\cite{EguiluzZimmermannCela,Zimmermann2004,Zimmermann2005,PachecoTraulsen,Pacheco2006a,Fu2008,Fu2009,Suzuki2008,Szolnoki2008,Poncela2009,Poncela2009a,SzolnokiPerc,VanSegbroeckSantos}, 
the iterated prisoner's dilemma\cite{EbelBornholdt}, 
and the snowdrift game 
\cite{PachecoTraulsen,Pacheco2006a,ZschalerTraulsenGross}
on adaptive networks. 
They showed numerically and analytically that a significantly increased level of cooperation can be achieved if individuals are able rewire their links \cite{EguiluzZimmermannCela,Zimmermann2004,Zimmermann2005,Fu2008,Fu2009,Suzuki2008,ZschalerTraulsenGross,Biely2007}
if links are formed and broken \cite{PachecoTraulsen,Pacheco2006a,Biely2007,Szolnoki2008,VanSegbroeckSantos,Szolnoki2009} 
or if new agents are added to the network \cite{Poncela2009,Poncela2009a}.  
Moreover, it has been shown that the adaptive interplay between the agents' strategies and the network topology
can lead to the emergence of distinguished agents from an initially homogeneous population \cite{EguiluzZimmermannCela,Zimmermann2000, Zimmermann2004,Zimmermann2005}. 

While important progress has been made in the investigation of games on adaptive networks, it is mostly limited to discrete networks, in which the agents can only assume a small number of different states, say, unconditional cooperation with all neighbours and unconditional defection. By contrast, continuous adaptive networks have received considerably less attention \cite{BalaGoyal,Schweitzer,Tomassini2009}. Most current models therefore neglect the ability of intelligent agents to maintain different levels of cooperation with different self-chosen partners \cite{Tomassini2009}.   

In this paper we propose a weighted and directed adaptive network model in which agents continuously and selectively reinforce advantageous collaborations.  
After a brief description of the model, we show in Sec.~\ref{SecCoordination} that the network generally approaches a state in which all agents make the same total cooperative investment and every reciprocated investment yields the same benefit. Despite the emergence of this high degree of coordination, the evolved networks are far from homogeneous. Typically the agents distribute their total investment heterogeneously among their collaborations, and each collaborations receives different investments from the partners. In Sec.~\ref{SecLeaders}, we show that this heterogeneity enables resource fluxes across the network, which allow agents holding distinguished topological positions to extract high payoffs. 
Thereafter, in Sec.~\ref{SecGiantComp}, we investigate further topological properties of the evolved networks and identify the transition in which large cooperating components are formed.
Finally, in Sec.~\ref{SecExploitation}, we focus on the appearance of unidirectional (unreciprocated) investments. Specifically, we identify three distinct scenarios in which unidirectional collaborations can arise and discuss their implications for the interaction topology. Our conclusions are summarized in Sec.~\ref{SecDiscussion}.    

\section{Model}\label{SecModel}
We consider a population of $N$ agents, representing for instance people, firms or nations, engaged in bilateral collaborative interactions. 
Each interaction is described by a continuous snowdrift game \cite{DoebeliHauert}, one of the fundamental models of game theory.
In this game, an agent $i$ can invest an amount of time/money/effort $e_{ij}\in\mathbb{R}_{0}^{+}$ into the collaboration with another agent $j$. 
Cooperative investments accrue equal benefits $B$ to both partners, but create a cost $C$ for the investing agent. 
Assuming that investments from both agents contribute additively to the creation of the benefit, the payoff received by agent $i$ from an interaction with an agent $j$ can then be written as 
\begin{equation}
P_{ij}=B\left(e_{ij}+e_{ji}\right)- C\left(e_{ij}\right). 
\end{equation}
The game thus describes the generic situation in which agents invest their personal resources to create a common good shared with the partner.

As an example of the snowdrift game, the reader may think of a scientific collaboration where two researchers invest their personal time in a project, while the benefit of the publication is shared between them. This example makes it is clear that the benefit of the collaboration must saturate when an extensive amount of effort is invested, whereas the cost to the an agents, measured for instance in terms of personal well-being, clearly grows superlinearly once the personal investment exceeds some hours per day.   
  
In the following we do not restrict the cost- and the benefit-functions, $B$ and $C$, to specific functional forms, except in the numerical investigations. However, we assume that both are differentiable and, moreover, that $B$ is sigmoidal and $C$ is superlinear (cf. Fig.~\ref{pocfig2}). These assumptions capture basic features of real-world systems such as inefficiency of small investments, saturation of benefits at high investments, as well as additional costs incurred by overexertion of personal resources and are widely used in the sociological and economic literature \cite{Oliver, Heckathorn}.

To account for multiple collaborations per agent, we assume that the benefits received from collaborations add linearly, whereas the costs are a function of the sum of investments made by an agent, such that the total payoff received by an agent $i$ is given by
\begin{equation}
P_i=\sum_{j\neq i}P_{ij}=\sum_{j\neq i}B\left(\sigma_{ij}\right)- C\left(\Sigma_i\right). 
\end{equation}
where $\Sigma_i:=\sum_{j=1}^N e_{ij}$ denotes the \emph{total investment} of the agent $i$ while $\sigma_{ij}:=e_{ij}+e_{ji}$ denotes the total investment made in the collaboration $ij$. 
This is motivated by considering that benefits from different collaborations, say different publications, are often obtained independently of each other, whereas the costs generated by different collaborations stress the same pool of personal resources of an agent. 

Let us emphasize that we do not restrict the investment of an agent further. While investments cannot be negative, no upper limit on the investments is imposed. Furthermore, the agents are free to make different investments in collaborations with different partners. Thus, to optimize its payoff, an agent can reallocate investments among its potential partners as well as change the total amount of resources invested.
 
For specifying the dynamics of the network, we assume the agents to be selfish, trying to increase their total payoff $P_i$ by a downhill-gradient optimization 
\begin{equation}\label{timeevolution}
	{\rm \frac{d}{dt}}e_{ij}= \frac{\partial}{\partial e_{ij}} P_i. 
\end{equation}

Every agent can cooperate with every other agent. Thus, the network of potential collaborations is fully connected and the deterministic time-evolution of the model system is given by a system of $N(N-1)$ ordinary differential equations of the form of Eq.~\ref{timeevolution}. 
The network dynamics, considered in the following, is therefore only the shifting of link weights $e_{ij}$. Note however that already the weight dynamics constitutes a topological change. As will be shown in the following, the agents typically reduce their investment in the majority of potential collaborations to zero, so that a sparse and sometimes disconnected network of non-vanishing collaborations is formed. Therefore the terminology of graph theory is useful for characterizing the state that the system approaches. Below, we use the term \emph{link} to denote only those collaborations that receive a non-vanishing investment $\sigma_{ij}$. A link is said to be \emph{bidirectional} if non-vanishing investments are contributed by both connected agents, while it is said to be \emph{unidirectional} if one agent makes a non-vanishing investment without reciprocation by the partner. Likewise, we use the term \emph{neighbours} to denote those agents that are connected to a focal agent by non-vanishing collaborations and the term \emph{degree} to denote the number of non-vanishing collaborations in which a focal agent participates.      

In the following, the properties of the model are investigated mostly by analytical computations that do not require further specifications. Only for the purpose of verification and illustration we resort to numerical integration of the ODE system. For these we use the functions
\begin{equation*}
B\left(\sigma_{ij}\right)= \frac{2\rho}{\sqrt{\vphantom{\left(\sigma_{ij}-\rho\right)^2}\tau+\rho^2}}+\frac{2(\sigma_{ij}-\rho)}{\sqrt{\tau+\left(\sigma_{ij}-\rho\right)^2}}\ ,\quad C\left(\Sigma_i\right)\,= \mu\left(\Sigma_i\right)^2.
\end{equation*}
For studying the time-evolution of exemplary model realizations by numerical integration, all variables $e_{ij}$ are assigned random initial values drawn independently from a Gaussian distribution with expectation value $1$ and standard deviation $10^{-14}$ constituting a homogeneous state plus small fluctuations. The system of differential equations is then integrated using Euler's method with variable step size $h$. In every timestep, $h$ is chosen such that no variable is reduced by more than half of its value in the step. If in a given timestep a variable $e_{ij}$ falls below a threshold $\epsilon<<1$ and the corresponding time derivative is negative, then ${\rm d}e_{ij}/{\rm dt}$ is set to zero for one step to avoid very small time steps. We emphasize that introducing the threshold $\epsilon$ is done purely to speed up numerical integration and does not affect the results or their interpretation.
In particular, we confirmed numerically that, the exact value of $\epsilon$ does not influence the final configuration that is approached. In all numerical results shown below $\epsilon=10^{-5}$ was used. 

\begin{figure}
\begin{minipage}{0.4\textwidth}
\begin{center}
\includegraphics[width=\textwidth]{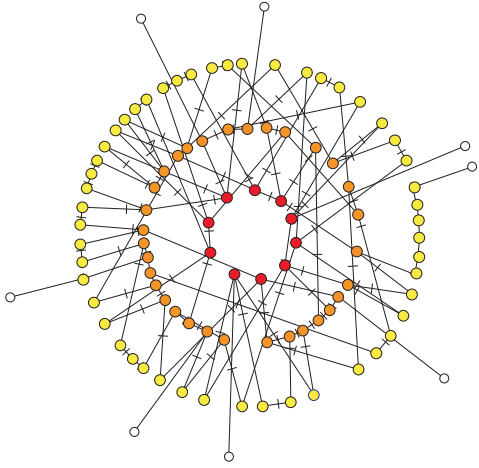}
\end{center}
\end{minipage}
\begin{minipage}{0.6\textwidth}
\sffamily{\caption{Network of collaborations in the final state. The nodes represent agents, links correspond to collaborations receiving non-vanishing investments $\sigma_{ij}$. The small dash on every link $ij$ is a fairness indicator: the further it is shifted toward one agent $i$, the lower the fraction, $e_{ij}/\sigma_{ij}$, of the investment agent $i$ contributes to the link. Agents extracting more payoff are shown in darker colour and are placed toward the center of the community. The size of a dot indicates the agents total investment $\Sigma_i$. In the final configuration the network exhibits a high degree of heterogeneity. Nevertheless all agents make the same total investment and all collaborations receive the same total investment. (Parameters: $\rho=0.65$, $\tau=0.1$, $\mu=1.5$)\label{pocfig1}}}
\end{minipage}     
\end{figure}

\section{Coordination of investments}\label{SecCoordination}
The numerical exploration of the system reveals frustrated, glass-like behavior; starting from a homogeneous configuration as described above, it approaches either one of a large number of different final configurations, which are local maxima of the total payoff.

\begin{figure}
\begin{flushright}
\includegraphics[width=0.8\textwidth]{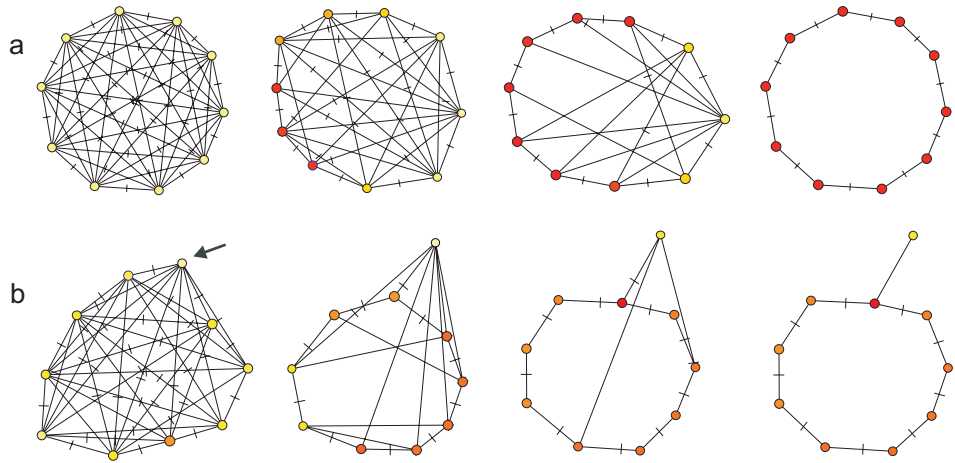}
\end{flushright}
\begin{center}
\sffamily{\caption{Time evolution of example networks from a homogeneous state. The different frames show snapshots of the network of collaborations at different times. a) In small systems the network sometimes self-organizes to homogeneous topologies in which all players extract the same payoff. b) If a player (arrow) tries to maintain too many links at too low investment, his partners will cease reciprocating investments, leading sometimes to unidirectional links.\label{pocfig1a}}}
\end{center}
\end{figure}

A representative example of an evolved network, and snapshots from the time-evolution of two smaller example networks are shown in Figs.~\ref{pocfig1},\ref{pocfig1a}, respectively. In the example networks only those links are shown that receive a non-vanishing (i.e. above-threshold) investment. Most of these non-vanishing links are \emph{bidirectional}, receiving investments from both of the agents they connect. Only rarely, \emph{unidirectional} links appear, which are maintained by one agent without reciprocation by the partner.    

For further investigations it useful to define a \emph{bidirectionally connected component} (BCC) as a set of agents and the bidirectional links connecting them, such that, starting from one agent in the set, every other agent in the set can be reached by following a sequence of bidirectional links. In the numerical investigations we observe that all bidirectional links within a BCC receive the same total investment in the final state. However, the investment $\sigma_{ij}$ made in every given link is in general not split equally among the two connected agents. Furthermore, all agents within a BCC make the same total cooperative investment $\Sigma_i$ in the final state. However, the investments $e_{ij}$ of one agent in different collaborations are in general different. The \emph{coordination} of total investments $\sigma_{ij}$, $\Sigma_i$ therefore arises although no agent has sufficient information to compute the total investment made by any other agent. 

We emphasize that the level of investments, which the agents approach is not set rigidly by external constraints but instead depends on the topology of the network of collaborations that is formed dynamically. This is evident for instance in differences of up to 20 \% between the level of investment that is reached in different BCCs of the same network. 

To understand how coordination of investment arises, we now formalize the observations made above. We claim that in our model in the final state the following holds:
Within a BCC (i) every agent makes the same total investment, and (ii) either all bidirectional links receive the same total investment or there are exactly two different levels of total investment received by bidirectional links. For reasons described below, the case of two different levels of total investment per link is only very rarely encountered. In this case every agent can have at most one bidirectional link that is maintained at the lower level of investment.  

\begin{figure}
\begin{flushright}
\includegraphics[width=0.85\textwidth]{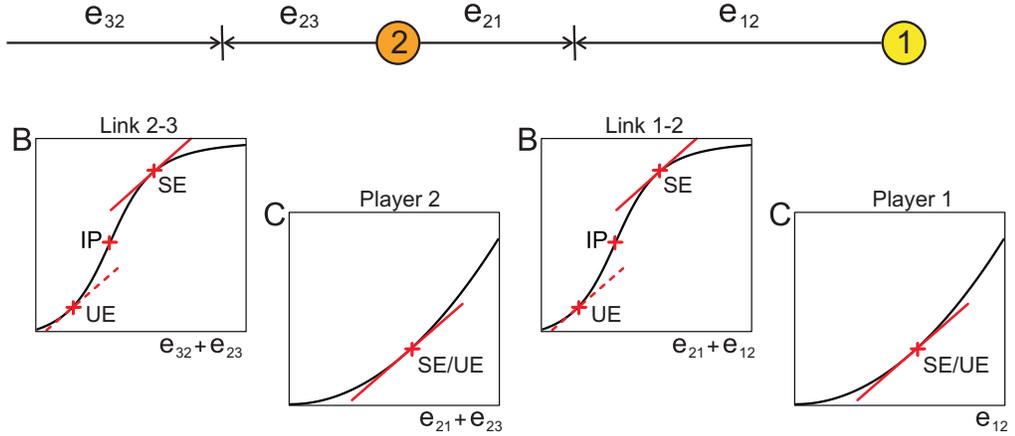}
\sffamily{\caption{Adjustment of investments. Shown are the perceived cost functions $C$ and benefit functions $B$ (insets) for the example of an agent 1 of degree one interacting with an agent 2 of degree two (sketched). The function $B$ depends on the sum of both agents' investments into the interaction while $C$ depends on the sum of all investments of one agent. 
In every equilibrium (SE or UE) stationarity demands that the slope of these functions is identical. This requires that the agents make identical total investments. In stable equilibria (SE), the operating point lies in general above the inflection point (IP) of $B$, whereas equilibria found below the IP are in general unstable (UE).
Therefore, in a stable equilibrium both links produce the same benefit and both agents make the same total investment. \label{pocfig2}}}
\end{flushright}
\end{figure}

We first focus on property (i). This property is a direct consequence of the stationarity of the final state. Consider a single link $ij$. Since both investments, $e_{ij}$ and $e_{ji}$, enter symmetrically into $\sigma_{ij}$, the derivative of the benefit with respect to either investment is $\partial B(\sigma_{ij}) / \partial e_{ij}=\partial B(\sigma_{ji}) / \partial e_{ji}=:B'(\sigma_{ij})$. Thus, if $e_{ij},e_{ji}>0$, the stationarity conditions ${\rm d}e_{ij}/{\rm dt}={\rm d}e_{ji}/{\rm dt}=0$ require
\begin{equation}\label{stationarity_cond}
	\frac{\partial}{\partial e_{ij}}C\left(\Sigma_{i}\right)=B'\!\left(\sigma_{ij}\right)=\frac{\partial}{\partial e_{ji}}C\left(\Sigma_{j}\right).  
	\end{equation}
This stipulates that the slope of the cost of the two interacting agents must match the slope of the shared benefit in the stationary state (Fig.~\ref{pocfig2}). Due to the symmetry of $\Sigma_{i}$, $\partial C(\Sigma_{i})/ \partial e_{ij}=\partial C(\Sigma_i)/\partial e_{ik}=:C'(\Sigma_{i})$ holds for all $i,j,k$. Therefore, 
Eq.~\eref{stationarity_cond} implies $C'(\Sigma_{i})=C'(\Sigma_{j})$. As we assumed $C$ to be superlinear, $C'$ is injective and it follows that $\Sigma_{i}=\Sigma_{j}=:\Sigma$, such that $i$ and $j$, are at a point of identical total investment. Iterating this argument along a sequence of bidirectional links yields (i). 

Let us remark that the stationarity of vanishing investments may be fixed due to the external constraint that investments have to remain non-negative. The stationarity condition for vanishing and uni-directional links, analogous to Eq.~\eref{stationarity_cond}, is therefore  
\begin{equation}\label{stat2}
	\frac{\partial}{\partial e_{ij}}C\left(\Sigma_{i}\right)\geq B'\!\left(\sigma_{ij}\right)\leq \frac{\partial}{\partial e_{ji}}C\left(\Sigma_{j}\right).  
\end{equation} 
Because of the inequalities that appear in this equation, the argument given above does not restrict the levels of total investment found in different components. For similar reasons agents that are only connected by unidirectional links can sustain different levels of investment, which is discussed in Sec.~\ref{SecExploitation}. 

We note that, although the network of potential interactions is fully connected, no information is transfered along vanishing links. Therefore, the equation of motion, Eq.~\ref{timeevolution}, should be considered as a local update rule, in the sense that it only depends on the state of the focal agent and on investments received from a small number of direct neighbours.  

In order to understand property (ii) we consider multiple links connecting to a single agent $i$. In an equilibrium the investment into each of the links has to be such that the slope of the benefit function of each link is identical. Otherwise, the payoff could be increased by shifting investments from one link to the other. Since the benefit function is sigmoidal, a given slope can be found in at most two points along the curve: one above and one below the inflection point (IP). By iteration, this implies that if a stationary level of investment is observed in one link, then the investment of all other links of the same BCC is restricted to one of two values, which amounts to the first sentence of (ii). 

For understanding why the case of two different levels of investments is rarely encountered the stability of steady states has to be taken into account. A local stability analysis, based on linearisation and subsequent application of Jacobi's signature criterion, is presented in the appendix. 
We show that for a pair of agents $ij$ connected by a bidirectional link, stability requires   
\begin{equation}\label{stab_cond1}
C''(\Sigma_i)>0\ \wedge \ 2B''(\sigma_{ij})-C''(\Sigma_i)<0, 
\end{equation} 
and every pair of links $ij$ and $ik$ connecting to the same agent $i$ has to satisfy 
\begin{equation}\label{stab_cond2}
	B''(\sigma_{ik})B''(\sigma_{ij})>\underbrace{C''(\Sigma_i)}_{>0}\left(B''(\sigma_{ik})+B''(\sigma_{ij})\right).
\end{equation}
Note that Eq.~\eref{stab_cond1} does not stipulate the sign of $B''(\sigma_{ij})$ as it only implies $2B''(\sigma_{ij})<C''(\Sigma_i)>0$. 
As Eq.~\eref{stab_cond1} applies also to the link $ik$, the same holds for $B''(\sigma_{ik})$. We therefore have to consider three different cases
when testing the compatibility of Eq.~\eref{stab_cond2} with  Eq.~\eref{stab_cond1}:
\begin{description}
	\item[a)] $B''(\sigma_{ik})<0$ and $B''(\sigma_{ij})<0$, (both investments above the IP)
	\item[b)] $B''(\sigma_{ik})>0$ and $B''(\sigma_{ij})>0$, (both investments below the IP)
	\item[c)] $B''(\sigma_{ik})>0$ and $B''(\sigma_{ij})<0$  (one investment above and one below the IP).
\end{description}
In case a), Eq.~\eref{stab_cond2} is trivially fulfilled as the left hand side has positive and the right hand side negative sign. 
In case b), Eq.~\eref{stab_cond2} and  Eq.~\eref{stab_cond1} are incompatible:
estimating the lower bound of the right hand side of \eref{stab_cond2} using the relation $C''(\Sigma)>2B''(\sigma_{ij})$ leads to the contradiction
\begin{eqnarray}\hspace{-2cm}
\overbrace{B''(\sigma_{ik})B''(\sigma_{ij})}^{:=X>0}&>C''(\Sigma_i)&\left(B''(\sigma_{ik})+B''(\sigma_{ij})\right) \nonumber\\											&>2B''(\sigma_{ij})&\left(B''(\sigma_{ik})+B''(\sigma_{ij})\right)=\underbrace{2B''(\sigma_{ij})B''(\sigma_{ik})}_{=2X}+\underbrace{2\left(B''(\sigma_{ij})\right)^2}_{>0}.\nonumber
\end{eqnarray}   
This shows that in a stable stationary state, every agent can at most have one link receiving investments below the IP. 
In case c), Eq.~\eref{stab_cond2} can in principle be satisfied. However, the equation still imposes a rather strong restriction on a positive $B''\left(\sigma_{ik}\right)$ requiring high curvature of the benefit function close to saturation. The restriction becomes stronger, when the degree of agent $i$ increases \cite{remark}. 

Bilateral links with investments below the IP can be excluded entirely, if the benefit function approaches saturation softly, so that the curvature above the inflection point remains lower or equal than the maximum curvature below the inflection point. For such functions, every pair $\sigma_{ik}<\sigma_{ij}$ of solutions to the stationarity condition $B'\left(\sigma_{ij}\right)=B'\left(\sigma_{ik}\right)=C'\left(\Sigma_i\right)$ yields a pair of coefficients $B''\left(\sigma_{ik}\right)>0, \ B''\left(\sigma_{ij}\right)<0$ violating \eref{stab_cond2}. In this case only configurations in which all links receive investments above the IP can be stable and hence all links produce the same benefit in the stable stationary states. This explains why the case of two different levels of cooperation is generally not observed in numerical investigations if realistic cost and benefit functions are used.

For understanding the central role the IP plays for stability consider that in the IP the slope of $B$ is maximal. Therefore, links close to the IP make attractive targets for investments. If the total investment into one link is below the IP then some disturbance raising (lowering) the investment increases (decreases) the slope, thus making the link more (less) attractive for investments. Hence, below the IP, a withdrawal of resources by one of the partners, no matter how slight, will make the collaboration less attractive, causing a withdrawal by the other partner and thereby launching the interaction into a downward spiral. Conversely, for links above the IP the gradual withdrawal of resources by one partner increases the attractiveness of the collaboration and is therefore compensated by increased investment from the other partner. In psychology both responses to withdrawal from a relationship are well known \cite{Baxter}. The proposed model can therefore provide a rational for their observation that does not require explicit reference to long term memory, planning, or irrational emotional attachment.        

For our further analysis property (ii) is useful as it implies that, although our model is in essence a dynamical system, the BCCs found in the steady states of this system can be analyzed with the tools of graph theory for undirected graphs. In the Secs.~\ref{SecLeaders}, \ref{SecGiantComp} we go one step further and treat not only the BCC but the whole network as an undirected graph. 
We thereby ignore the differences between directed and undirected links in order to study properties such as the degree- and component-size distributions before we continue in Sec.~\ref{SecExploitation} with a more detailed investigation of directed links and their topological implications.

\section{Distinguished topological positions}\label{SecLeaders}
Despite the coordination described above, the payoff extracted by agents in the final state can differ significantly. This is remarkable because the agents follow identical rules and the network of collaborations is initially almost homogeneous with respect to degree, link weights, and neighbourhood.

Because all bidirectional links in a BCC produce the same benefit, the total benefit an agent receives is proportional to the degree of the agent. By contrast, the cost incurred by an agent does not scale with the degree, but is identical for all agents in the BCC, because agents of high degree invest a proportionally smaller amount into their collaborations. Topological positions of high degree thus allow agents to extract significantly higher benefits without requiring more investment.    

\begin{figure}
\begin{minipage}{0.45\textwidth}
\begin{center}
\includegraphics[width=0.9\textwidth]{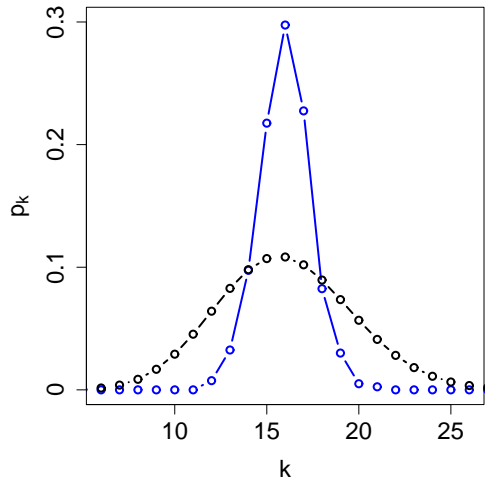}
\end{center}
\end{minipage}
\begin{minipage}{0.55\textwidth}
\vspace{-2.0cm}
\sffamily{\caption{Degree heterogeneity in self-organized networks of collaborations. In comparison to a random graph (black), the degree distribution of the evolved networks is relatively narrow (blue). Parameters are chosen to obtain networks with identical mean degree. Results are averaged over 100 networks of size $N=100$.\label{pocfig3}}}
\end{minipage}
\end{figure}

\begin{figure}[b]
\begin{minipage}{0.45\textwidth}
\begin{center}
\includegraphics[width=0.6\textwidth]{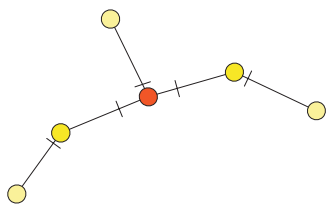}
\end{center}
\end{minipage}
\begin{minipage}{0.55\textwidth}
\vspace{-0.5cm}
\sffamily{\caption{Redistribution of investments. Even in small networks investments flow toward agents of high connectivity. This flow is apparent in the position of the fairness indicators on the links, cf.~Fig.~\ref{pocfig1}, caption. \label{pocfig4}}}
\end{minipage}      
\end{figure} 
The payoff distribution in the population is governed by the degree distribution $p_k$ describing the relative frequency of agents with degree $k$. Figure~\ref{pocfig3} shows a representative degree distribution of an evolved networks in the final state. While the finite width of the distribution indicates heterogeneity, the distribution is narrower, and therefore fairer, than that of an Erd{\H{o}}s-R{\'e}nyi random graph, which constitutes a null-model for randomly assembled network topologies. We verified that variance of the evolved network is below the variance of a random graph for the whole range of admissible mean degree $\bar{k}$ in a network of given size. 

Although the snowdrift game is not a zero-sum game, payoffs cannot be generated arbitrarily. In order to sustain the extraction of high payoffs by agents of high degree, investments have to be redistributed across the network. In the definition of our model, we did not include the transport of resources directly. Nevertheless, a redistribution of investments arises indirectly from the asymmetry of the agents' investments. This is illustrated in Fig.~\ref{pocfig4}. Consider for instance an agent of degree 1. This agent necessarily focuses his entire investment on a single collaboration. Therefore, the partner participating in this collaboration only needs to make a small investment to make the collaboration profitable. He is thus free to invest a large portion of his total investment into links to other agents of possibly higher degree. In this way investments flow toward the regions of high degree where high payoffs are extracted.

\section{Formation of large components}\label{SecGiantComp}
To explore the topological properties of the networks of collaborations in the final state further, we performed an extensive series of numerical integrations runs in which we varied all parameters in a wide range. These revealed that an important determinant of the topology is the mean degree $\bar{k}=2L/N$, where $L$ denotes the number of links and $N$ the number of agents in the network. Given two evolved networks with similar $\bar{k}$, one finds that the networks are also similar in other properties such as the component-size distribution, clustering coefficient, and the fraction of collaborations that are unidirectional. We therefore discuss the topological properties of the evolved networks as a function of $\bar{k}$, instead of the original model parameters. 

\begin{figure}
\begin{flushright}
\includegraphics[width=0.9\textwidth]{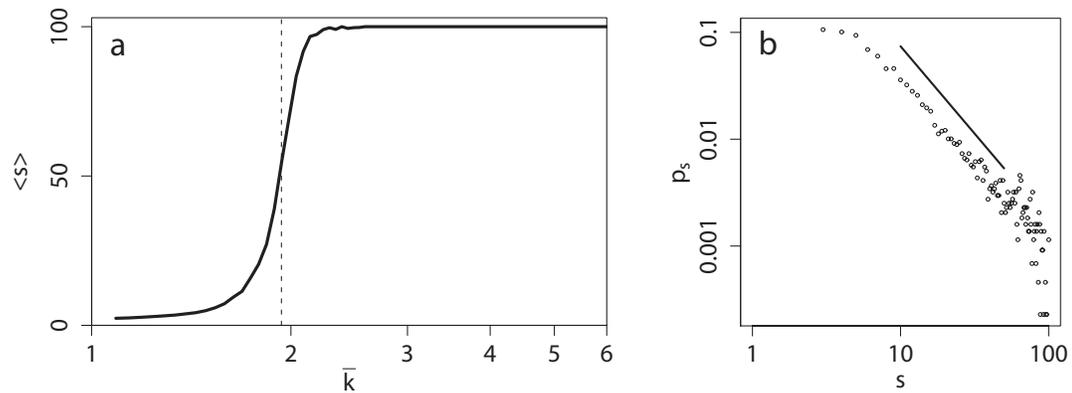}
\end{flushright}
\begin{center}
\sffamily{\caption{Giant component transition. (a) At $\bar{k}=1.91$ the expected size $\langle s\rangle$ of a network component changes from $O(1)$ to $O(N)$. (b) Even in the relatively small networks of 100 nodes a power-law shape starts to appear in the component-size distribution obtained from the final states of 750 network realizations with a mean degree a mean degree $\bar{k}= 1.91$. \label{pocfig5}}}
\end{center}
\end{figure}

We first consider the expected size $\langle s\rangle$ of a network component to which a randomly chosen agent belongs. In contrast to the BCC's discussed above, unidirectional collaborations are now taken into account in the computation of component sizes. The value of $\langle s\rangle$ in the evolved network as a function of $\bar{k}$ is shown in Fig.~\ref{pocfig5}a. The figure reveals that large components begin to appear slightly below $\bar{k}=2$. Because of the difficulties related to integrating $N(N-1)$ differential equations, our numerical investigations are limited to networks of up to 100 agents. While it is therefore debatable whether the observed behaviour qualifies as a phase transition, it can be related to the giant component transition commonly observed in larger networks. 

In the giant component transition a component is formed that scales linearly with network size. In the absence of higher correlations the transition occurs at $\bar{q}=1$ \cite{Newman:Review}, where $\bar{q}$ is the mean excess degree of the network, i.e., the number of additional links found connected to a agent that is reached by following a random link. 

In Erd{\H{o}}s-R{\'e}nyi random graphs, $\bar{q}=\bar{k}$, therefore the giant component transition takes place at $\bar{k}=1$. 
In the present model the transition in $\langle s\rangle$ is shifted to higher values of $\bar{k}$ because of the nature of the underlying snowdrift game. The snowdrift game favors cooperation in the sense that for an agent of degree zero it is always advantageous to initiate an interaction. Therefore $\bar{k}=1$ is the lowest possible value that can be observed in evolved networks. Further, any evolved network with $\bar{k}=1$ invariably consists of isolated pairs, which precludes the existence of a giant component. Finally, the relatively narrow degree distribution of the evolved networks implies $\bar{q}<\bar{k}$ and therefore $\bar{k}>1$ at the transition. 

To estimate an upper limit for the connectivity at which the giant component transition occurs, it is useful to consider degree homogeneous networks. In these networks the degree distribution is a delta function and $\bar{q}=\bar{k}-1$, so that the transition occurs at $\bar{k}=2$. In the networks evolved in the proposed model we can therefore expect a critical value of $\bar{k}$ between one and two. Based on numerical results 
we estimate that the giant component transition in the present model occurs at $\bar{k}\approx 1.91$  (Fig.~\ref{pocfig5}). At this value a power-law distribution of component sizes, which is a hallmark of the giant-component transition, begins to show already in relative small networks with $N=100$.   

\section{Unreciprocated collaborative investments}\label{SecExploitation}
While in Sec.~\ref{SecCoordination} we have mainly considered bidirectional links, and in Sec.\ref{SecLeaders} and \ref{SecGiantComp} only distinguished between vanishing and non-vanishing links, we will now focus on unidirectional links, which one partner maintains without reciprocation by the other. 
The presence of such links in collaboration networks was recently discussed in detail by \cite{Schweitzer}. 

For the discussion below it is advantageous to consider the mean degree of agents in a connected component $\langle k \rangle=2l/n$, where $n$ and $l$ are the number of agents and links in the component.      
Note that in large components $\langle k \rangle \approx \bar{k}$ while the two properties can be significantly different in small components. 
In contrast to $\bar{k}$, $\langle k\rangle$ allows us to infer global topological properties: 
Components with $\langle k\rangle<2$ are trees. 
Components with $\langle k\rangle=2$ contain exactly one cycle to which trees might be attached. And, components with $\langle k\rangle>2$ contain more than one cycle, potentially with trees attached.
 
As in the previous section, the term component refers to maximal subgraphs which are connected by bidirectional and/or unidirectional links. According to this definition a component may, beside one or more BCCs, contain agents, which only have unidirectional links. In the following we denote the set of these agents as the non-BCC part of the component (nBCC). For the sake of simplicity we focus on components which contain only one BCC, but note that the case of multiple BCCs can be treated analogously.

Unlike the BCC, the nBCC is not a subcomponent but only a set of agents which are not necessarily connected.
Nevertheless, numerical results show that (i*) all nBCC agents make the same total investment $\Sigma_{\rm n}$ and (ii*) all unidirectional links maintained by nBCC agents receive the same total investment $\sigma_{\rm n}$. 
While property (ii*) can be understood analogously to property (ii) of BCCs, property (i*) cannot be ascribed to stationarity or stability conditions but seems to result from optimality restrictions.  As a consequence of 
the properties (i*) and (ii*) the number of outgoing links $m:=\Sigma_{\rm n}/\sigma_{\rm n}$ is identical for all agents in the nBCC.  
  
\begin{figure}
\begin{flushright}
\includegraphics[width=0.85\textwidth]{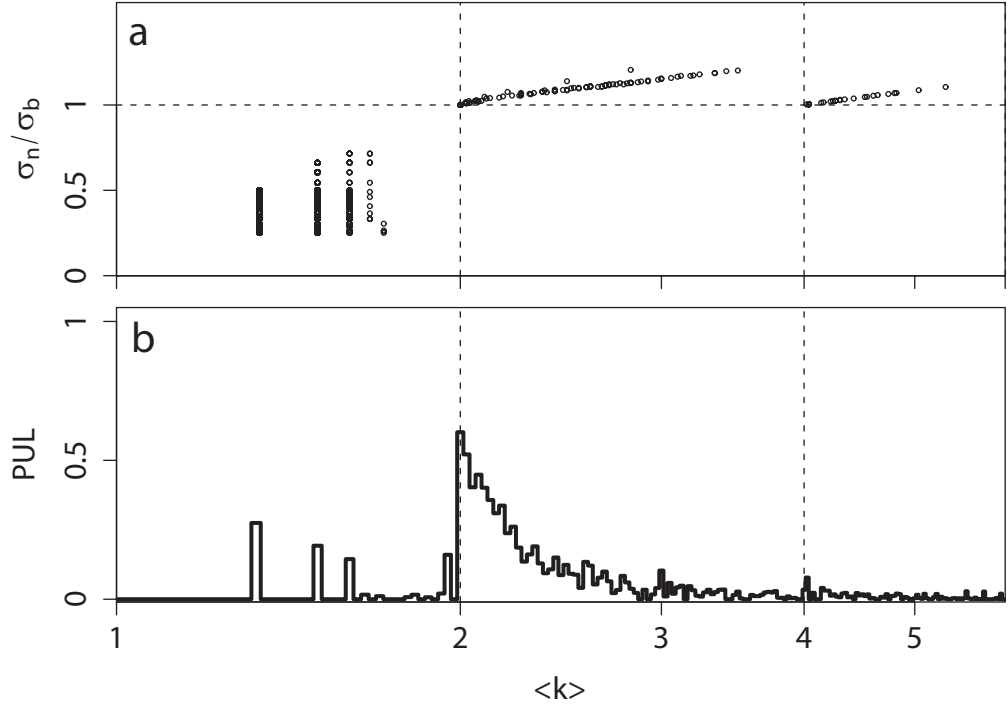}
\sffamily{\caption{Unidirectional investments and proportion of unidirectional links. 
(a) The ratio between the investment in unidirectional and the investment in bidirectional links from the same component, $\sigma_{\rm n}/\sigma_{\rm n}$, equals $1$ for $\langle k\rangle=2m,\ m\in \mathbb{N}$. $\sigma_{\rm n}/\sigma_{\rm b}>1$ applies to $\langle k\rangle>2\neq2m$, $\sigma_{\rm n}/\sigma_{\rm b}<1$ to $\langle k\rangle<2$. (b) For $\langle k\rangle<2$ the average proportion of unidirectional links (PUL) features discrete peaks. As every tree must have a bidirectional core, the smallest $\langle k\rangle$ with non-zero PUL is  $\langle k\rangle=4/3$. It corresponds to components with 3 agents and 2 links one of which can be unidirectional. \label{pocfig6}}}
\end{flushright}
\end{figure}

So far we have decomposed a component into the BCC and the nBCC. Within each subset, all agents make the same total investment, and all links receive the same total investment, therefore each subset can be characterized by two parameters, $\Sigma_{\rm b},\ \sigma_{\rm b}$ for BCC and $\Sigma_{\rm n}, \ \sigma_{\rm n}$ for the nBCC. 
To recombine the subsets and infer properties of the whole component, we need to study the relation between these four parameters. 

The central question guiding our exploration is, why do agents not start to reciprocate the unidirectional investments. The lack of reciprocation implies that the unidirectional links are either 
less attractive or just as attractive as bidirectional links. 
We distinguish the two scenarios
\begin{description} 
\item[a)] $B'(\sigma_{\rm b})=B'(\sigma_{\rm n})$,  
\item[b)] $B'(\sigma_{\rm b})>B'(\sigma_{\rm n})$.
\end{description}
In case a) the unidirectional collaborations are as attractive as targets for investments as bidirectional collaborations. In typical networks, where all remaining links receive investments above the IP this implies $\sigma_{\rm b}=\sigma_{\rm n}=\sigma$. Furthermore, in case a) the stationarity condition, Eq.~\eref{stationarity_cond}, requires that $C'(\Sigma_{\rm b})=C'(\Sigma_{\rm n})$, which stipulates $\Sigma_{\rm b}=\Sigma_{\rm n}=:\Sigma$. Therefore the whole component consists of agents making an investment $\Sigma$ and links receiving an investment $\sigma$.
 
Conservation of investments within a component implies $l\sigma=n\Sigma$ and hence 
\begin{equation}\label{meank}
	\langle k\rangle= 2\frac{l}{n}=2\frac{\Sigma}{\sigma}.
\end{equation}
We know further that $\Sigma/\sigma=\Sigma_{\rm n}/\sigma_{\rm n}=m\in \mathbb{N}$, where $m$ is the number of outgoing links of
an agent in the nBCC. 
Inserting $\Sigma/\sigma=m$ in Eq.\eref{meank} yields $\langle k\rangle=2m$, showing that unidirectional links that are as attractive as bidirectional links can only occur in components in which mean degree, $\langle k\rangle $, is an integer multiple of 2. This matches the numerical data displayed in Fig.~\ref{pocfig6}a, which shows that $\sigma_{\rm n}/\sigma_{\rm b}=1$ is observed in components with $\langle k\rangle=2$ and $\langle k\rangle=4$.  

\begin{figure}
\begin{flushright}
\includegraphics[width=0.85\textwidth]{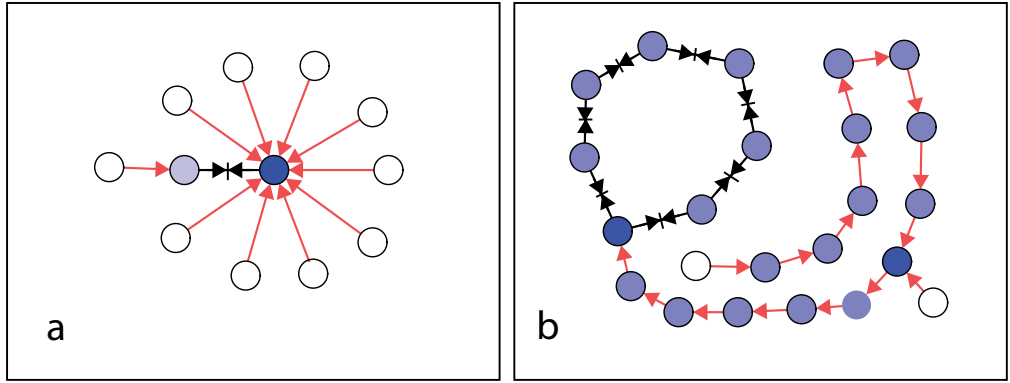}
\sffamily{\caption{Topological arrangement of unidirectional links (shown in red). (a) For $\langle k\rangle<2$, unidirectional links connect individual nBCC agents with a BCC core. (b) For $\langle k\rangle\geq2$ unidirectional links are arranged in long chains which is shown here for $\langle k\rangle=2$. For $\langle k\rangle>2$ typical components become too large to be presented in this way.  \label{pocfig7}}}
\end{flushright}\end{figure}

It is remarkable that observing $\sigma_{\rm n} = \sigma_{\rm b}$ in a pair of collaborations is sufficient to determine the mean degree of the whole component. Moreover components in which the mean degree is exactly 2 have to consist of a single cycle potentially with trees attached. In the numerical investigations we mostly observe cycles of bidirectional links to which trees of unidirectional links are attached, as shown in Fig.~\ref{pocfig7}b.

In case b) the bidirectional links are more attractive targets for investments than unidirectional links.
In typical networks with $\sigma_{\rm b},\sigma_{\rm n} \geq \sigma_{\rm IP}$ this implies $\sigma_{\rm b}<\sigma_{\rm n}$.
Now the stationarity condition, Eq.~\eref{stationarity_cond}, demands that $C'(\Sigma_{\rm b})>C'(\Sigma_{\rm n})$,
so that unidirectional links receive a higher investment than bidirectional links. By contrast the total investment made by an agent investing in bidirectional links is higher than the one made by agents investing in unidirectional links, i.e. 
\begin{equation}\label{kg2_eq}
	\sigma_{\rm b}<\sigma_{\rm n}\leq\Sigma_{\rm n}<\Sigma_{\rm b}.
\end{equation}
This relationship restricts the connectivity in the BCC to $\langle k\rangle_{BCC}:=2 \Sigma_{\rm b}/\sigma_{\rm b}>2$,
which implies $\langle k\rangle>2$, because the mean degree of the component cannot be smaller than 
2 if a subcomponent already has a degree greater than 2. Therefore, we find that unidirectional links that are less attractive than bidirectional links only occur in components in which the mean degree is larger than $2$, but not an integer multiple of 2 (cf. Fig.~\ref{pocfig6}a). As such links are only found at $\bar{k}$ beyond the giant component transition they occur typically in large components as shown in Fig.~\ref{pocfig1}.

In numerical investigations, we also observe some unidirectional links in components with $\langle k\rangle<2$ (cf. Fig.~~\ref{pocfig6}b).
To explain these we have to consider case b) but relax the assumption that both, $\sigma_{\rm n}$ and $\sigma_{\rm b}$ are above the IP. Thus, we obtain case c), about which we know that the unidirectional links are less attractive than bidirectional links, $\Sigma_{\rm n} < \Sigma_{\rm b}$, and that the unidirectional link only receives investments from one agent, i.e.,~$\sigma_{\rm n}\leq\Sigma_{\rm n}$. Moreover, $\langle k\rangle<2$ implies $\langle k\rangle_{BCC}<2$ and therefore $\Sigma_{\rm b} < \sigma_{\rm b}$. Therefore
\begin{equation}
\sigma_{\rm n}\leq\Sigma_{\rm n}<\Sigma_{\rm b}<\sigma_{\rm b},
\end{equation}
which shows that unidirectional links can only appear in components with $\langle k\rangle<2$ if the investment received by unidirectional links is smaller than the investment received by bidirectional links. Satisfying $\sigma_{\rm n}<\sigma_{\rm b}$ and $B'(\sigma_{\rm n})<B'(\sigma_{\rm b})$ simultaneously requires $\sigma_{\rm n}<\sigma_{\rm IP}$. The components with $\langle k\rangle<2$, in which such links are found, are trees formed by a core of bidirectional links, to which individual agents are attached by unidirectional links (Fig.~\ref{pocfig7}a). Chains of unidirectional links, as we have observed in case a), cannot appear for $\langle k\rangle<2$ as this would mean that some agents would have one incoming and one outgoing link below the IP, which is ruled out by a trivial extension of the reasoning from Sec.~\ref{SecCoordination}.
     

\section{Conclusions}\label{SecDiscussion}
In this paper we have proposed a model for the formation of complex collaboration networks between self-interested agents. In this model the evolving network is described by a large system of deterministic differential equations allowing agents to maintain different levels of cooperation with different partners. 

We showed analytically that bidirectionally communities are formed, in which every agent makes the same total investment and every collaboration provides the same benefit. In contrast to models for cooperation on discrete networks, the present model thereby exhibits a high degree of coordination which can be interpreted as a precursor of a social norm. We emphasized that coordination is generally achieved although single agents possess insufficient information for computing the total investment made by any other agent and although the level of cooperation that is reached in a community is not fixed rigidly by external constraints. 

Despite the high degree of coordination, we observed the appearance of privileged agents, reminiscent of the leaders emergind in \cite{Zimmermann2000}. In the model proposed in the present paper, the privileged agents hold distinguished topological positions of high degree centrality allowing them to extract much higher payoffs than other agents, while making the same cooperative investment. However, we found that in the absence of further mechanism reinforcing differences the assembled topologies were fairer than random graphs.   
  
Although our primary aim was to investigate the formation of social networks, some aspects of the behavior of social agents are reminiscent of results reported in psychology. For instance our investigation showed that agents can react to the withdrawal of investment by a partner either by mutual withdrawal of resources or by reinforcing the collaboration with increased investment. Our analysis provides a rational which links the expected response to the withdrawal of resources to an inflection point of an assumed benefit function.

Furthermore, we investigated under which conditions non-reciprocated collaborations appear. 
Here, our analysis revealed that such unidirectional collaborations can appear in three distinct scenarios, which can be linked to topological properties of the evolving networks. In particular exploited agents whose investments are not reciprocated invest less than the average amount of resources in their links when occurring in small components, but more than the average amount, when integrated in large components.  

We believe that the results from the proposed model can be verified in laboratory experiments in which humans interact via a computer network. Such experiments may confirm the topological properties of the self-organized networks reported here and may additionally provide insights into the perceived cost and benefit functions that humans attach to social interactions. 

Furthermore, results of the proposed model may be verified by comparison with data on collaboration networks between people, firms or nations. This comparison may necessitate modifications of the model to allow for instance for slightly different cost functions for the players. Most of these extensions are straight forward and should not alter the predictions of the model qualitatively. For instance in the case of heterogeneous cost functions, players will make different total investments, but will still approach an operating point in which the slope of their cost function is identical. Further, coordination should persist even if the network of potential collaborations is not fully connected. Finally, but perhaps most importantly our analytical results do not rely heavily on the assumption that only two agents participate in each collaboration. Most of the results can therefore be straight-forwardly extended to the case of multi-agent collaborations.   

Our analytical treatment suggests that the central assumption responsible for the emergence of coordination is that the benefit of a collaboration is shared between the collaborating agents, but is independent of their other collaborations, whereas the cost incurred by an agent's investment depends on the sum of all of an agent's investments. Because this assumption seems to hold in a relatively large range of applications we believe that also the emergence of coordination and leaders by the mechanisms described here should be observable in a wide range of systems. 

The analysis presented in this paper has profited greatly from the dual nature of the model, combining aspects of dynamical systems and complex network theory. In particular our analytical investigations were based on the application of Jacobi's signature criterion to the system's Jacobian matrix. 
Apart from the symmetry of the Jacobian, this `double-Jacobi' approach does not depend on specific features of model under consideration. The same approach can therefore be used to address significant extensions of the present model. We therefore believe that also beyond the field of social interactions, the double-Jacobi approach will prove to be a useful tool for the analytical exploration of the weighted adaptive networks that appear in many applications.     


\appendix

\section{Stability condition}
To determine the local asymptotic stability of the steady states we study the Jacobian matrix ${\rm \bf J}\in\mathbb{R}^{N(N-1)\times N(N-1)}$ defined by $J_{(ij)(kl)}=\partial\dot{e_{ij}}/\partial e_{kl}$.
The terms contained in this matrix can be grouped into three different types  
\begin{eqnarray}
A_{ij}:=&\ \frac{\partial\dot{e_{ij}}}{\partial e_{ij}}
= \frac{\partial^2}{(\partial e_{ij})^2}B\left(\sigma_{ij}\right) - \frac{\partial^2}{\left(\partial e_{ij}\right)^2}C\left(\Sigma_i\right)\label{Acodiff_eq}  \\
P_{ij}:=&\ \frac{\partial\dot{e_{ij}}}{\partial e_{ji}}=\ \frac{\partial}{\partial e_{ji}}\frac{\partial}{\partial e_{ij}}B\left(\sigma_{ij}\right)\\ 
K_{i}:=&\ \frac{\partial\dot{e_{ij}}}{\partial e_{il}} =\ -\frac{\partial}{\partial e_{il}}\frac{\partial}{\partial e_{ij}}C\left(\Sigma_i\right)\label{codiff_eq}
\end{eqnarray}
albeit evaluated at different points. For reasons of symmetry 
\begin{eqnarray}
\frac{\partial}{\partial e_{ji}}\frac{\partial}{\partial e_{ij}}\;B\left(\sigma_{ij}\right)=&\ \frac{\partial^2}{\left(\partial e_{ij}\right)^2}\;B\left(\sigma_{ij}\right)=:B''\left(\sigma_{ij}\right)\nonumber\\
\frac{\partial}{\partial e_{il}}\frac{\partial}{\partial e_{ij}}\;C\left(\Sigma_i\right)=&\ \frac{\partial^2}{\left(\partial e_{ij}\right)^2}\;C\left(\Sigma_i\right)=:C''\;\left(\Sigma_i\right)\ , \nonumber
\end{eqnarray}
and consequentially $P_{ij}=P_{ji}$, and $A_{ij}=P_{ij}+K_{i}$. Ordering the variables according to the mapping $M:\mathbb{N}\times\mathbb{N}\rightarrow\mathbb{N};\ (i,j)\rightarrow N(i-1)+j$ the Jacobian can be written in the form
\begin{equation*}{\rm \bf J}=\left(\begin{tabular}{cccccccccccc}
		\textcolor{red}{$A_{12}$}&$K_1$&\textcolor{red}{$P_{12}$}&0&0&0\\
		$K_1$&$A_{13}$&0&0&$P_{13}$&0\\
		\textcolor{red}{$P_{12}$}&0&\textcolor{red}{$A_{21}$}&$K_2$&0&0\\
		0&0&$K_2$&$A_{23}$&0&$P_{23}$\\
		0&$P_{13}$&0&0&$A_{31}$&$K_3$\\
		0&0&0&$P_{23}$&$K_3$&$A_{32}$\\
		\end{tabular}\right)\ ,\end{equation*}
which is shown here for $N=3$. As each cooperation $ij$ is determined by a pair of variables $\left(e_{ij}, e_{ji}\right)$, each $P_{ij}$ occurs twice forming quadratic subunits with the corresponding entries $A_{ij}$ and $A_{ji}$.
Subsequently, we restrict ourselves to the submatrix ${\rm \bf J^s}$ of ${\rm \bf J}$, which only captures variables $e_{ij}$ belonging to  `non-vanishing' links. 
As argued before, `vanishing links', i.e.~links with $\sigma_{ij}=0$, are subject to stationarity condition \eref{stat2}. 
If $C'\left(\Sigma_{i}\right)>B'\left(0\right)$, their stability is due to the boundary condition $e_{ij}\geq0$ and is independent of the second derivatives of $C$ and $B$. Hence, they can be omitted from the subsequent analysis.
This means in particular that the spectra of different topological components of the network decouple and can thus be treated independently. 

All eigenvalues of the real, symmetric matrix ${\rm \bf J^s}$ are real.
According to Jacobi's signature criterion the number of negative eigenvalues equals the number of changes of sign in the sequence $1,D_{1},\ldots,D_{r}$ where $r$ is the rank of ${\rm \bf J^s}$ and $D_{q}:=\det\left(J^{s}_{ik}\right)$, $i,k=1,\ldots,q$ \cite{Zeidler}. 
In a stable system the sequence has to alternate in every step. A necessary condition for stability is therefore alternation in the first steps $1, D_{1}, D_{2}$.

By means of an even number of column and row interchanges the above stated form of ${\rm \bf J^s}$ can always be transformed such that the first $2\times2$ block reads \begin{equation*}
\left(\begin{tabular}{cc}
$A_{ij}$&$P_{ij}$\\
$P_{ij}$&$A_{ji}$\\
\end{tabular}\right).
\end{equation*} 
Since we assume that $ij$ is a non-vanishing link, and, hence, $i$ and $j$ to be in the same component, both agents make the same total investment $\Sigma$. It follows from definition \eref{codiff_eq} that $K_i=K_j=:K$ and therewith that $A_{ij}=A_{ji}$. 
Thus, the sequence $1, D_{1}, D_{2}$ alternates if 
\begin{eqnarray}
D_{1}&=P_{ij}+K<0\quad \wedge \label{cond1mod_eq}\\
D_{2}&=\left(2P_{ij}+K\right)K>0.\label{cond2mod_eq}
\end{eqnarray}
Equation \eref{cond2mod_eq} stipulates that $K$ and $\left(2P_{ij}+K\right)$ have the same sign. 
Of the two possible scenarios
	\begin{equation} \left(2P_{ij}+K\right),K<0 \quad \mbox{and} \quad \left(2P_{ij}+K\right),K>0
\end{equation}
the second is ruled out by Eq.~\eref{cond1mod_eq}: If $K>0$, it follows from Eq.~\eref{cond1mod_eq} that $P_{ij}<-K<0$, which contradicts $\left(2P_{ij}+K\right)>0$. 
Hence, the necessary conditions for stability, Eqs.~\eref{cond1mod_eq}, \eref{cond2mod_eq}, require
\begin{equation} \label{scen1_eq}
	K<0\quad \wedge \quad \left(2P_{ij}+K\right)<0 \ .
\end{equation}

If either agent $i$ or agent $j$ has another bilateral link, say $ik$, it is furthermore possible to transform ${\rm \bf J^s}$ by an even number of row and line interchanges such that the first $2\times2$ block reads 
\begin{equation}\label{repr2_eq}
\left(\begin{tabular}{cc}
$A_{ij}$&$K$\\
$K$&$A_{ik}$\\
\end{tabular}\right).
\end{equation} 
In this representation the sequence $1, D_{1}, D_{2}$ alternates if 
\begin{eqnarray}
D_{1}=&A_{ij}=P_{ij}+K<0 \label{cond3_eq}\\
D_{2}=&P_{ik}P_{ij}+\left(P_{ik}+P_{ij}\right)K>0.\label{cond4_eq}
\end{eqnarray} 
Condition \eref{cond4_eq} can then be written as 
\begin{equation}\label{cond4mod_eq}
P_{ik}P_{ij}>-K\ \left(P_{ik}+P_{ij}\right). 
\end{equation}
Inserting the definitions Eqs.\eref{Acodiff_eq}-\eref{codiff_eq} in Eqs.~\eref{scen1_eq} and \eref{cond4mod_eq} yields the stability conditions cited in the main text as Eqs.~\eref{stab_cond1}-\eref{stab_cond2}. 

\noappendix


\begin{thebibliography}{}
\bibitem{remark} This can be shown by taking determinants $D_q$ with $q>2$ into account.
\end{thebibliography}
\end{document}